\documentclass[journal]{IEEEtran}
\usepackage{cite}
\usepackage{amsmath,amssymb,amsfonts}
\usepackage{algorithm}
\usepackage{algpseudocode}
\usepackage{graphicx}
\usepackage{textcomp}
\usepackage[export]{adjustbox}
\usepackage{xcolor}
\usepackage{gensymb}
\usepackage{listings}
\usepackage{url}

\usepackage{multirow}
\usepackage{footnote}

\usepackage[margins]{trackchanges}
\ifCLASSINFOpdf
\else
\fi
\usepackage[margins]{trackchanges} 

\begin{document}
%
\title{Fully Automatic In-Situ Reconfiguration of RF Photonic Filters in a CMOS-Compatible Silicon Photonic Process}
%
%
%

\author{Md~Jubayer Shawon,~\IEEEmembership{Student Member,~IEEE,}
        and~Vishal~Saxena,~\IEEEmembership{Senior Member,~IEEE}
\thanks{The authors are with the Department of Electrical and Computer Engineering, University of Delaware, Newark,
DE, 19716 USA e-mail: shawon@udel.edu.}
}

%
%

\markboth{This is an author-produced, under peer-review version (IEEE Journal). IEEE Copyright restrictions may apply.}%
{Shell \MakeLowercase{\textit{et al.}}: Bare Demo of IEEEtran.cls for IEEE Journals}
%



\maketitle
\pagestyle{empty}

\begin{abstract}
Automatic reconfiguration of optical filters is the key to novel flexible RF photonic receivers and Software Defined Radios (SDRs). Although silicon photonics (SiP) is a promising technology platform to realize such receivers, process variations and lack of in-situ tuning capability limits the adoption of SiP filters in widely-tunable RF photonic receivers. To address this issue, this work presents a first `in-situ' automatic reconfiguration algorithm and demonstrates a software configurable integrated optical filter that can be reconfigured on-the-fly based on user specifications. The presented reconfiguration scheme avoids the use of expensive and bulky equipment such as Optical Vector Network Analyzer (OVNA), does not use simulation data for reconfiguration, reduces the total number of thermo-optic tuning elements required and eliminates several time consuming configuration steps as in the prior art. This makes this filter ideal in a real world scenario where user specifies the filter center frequency, bandwidth, required rejection $\&$ filter type (Butterworth, Chebyshev, etc.) and the filter is automatically configured regardless of process, voltage $\&$ temperature (PVT) variations. We fabricated our design in AIM Photonics' Active SiP process and have demonstrated our reconfiguration algorithm for a second-order filter with 3dB bandwidth of 3 GHz, 2.2 dB insertion loss and $>$30 dB out-of-band rejection using only two reference laser wavelength settings. Since the filter photonic integrated circuit (PIC) is fabricated using a CMOS-compatible SiP foundry, the design is manufacturable with repeatable and scalable performance suited for its integration with electronics to realize complex chip-scale RF photonic systems.
\end{abstract}


%
\IEEEpeerreviewmaketitle

\section{Introduction}
%
%
%
%
\IEEEPARstart{I}{ntegrated} radio-frequency (RF) photonics is rapidly emerging as a technology enabler of demanding application scenarios which require capabilities beyond those of traditional electronic systems. These capabilities include ultra-wide bandwidth, exceptional low latency, long-distance routing and immunity to electromagnetic interference (EMI) \cite{capmany2013microwave, urick2015fundamentals, yi2017integrated, minasian2013microwave, marpaung2013integrated, zhang2015silicon}. For any RF photonic integrated circuit (\textbf{IC}), optical filters are essential building blocks. The desirable feature that distinguishes RF photonic filters from their electronic, microwave and micro-electromechanical systems (MEMS) counterparts is their tunability over a very wide range of center frequencies (1 to 100s of GHz) and very wide bandwidth. Consequently, RF photonic filters offer unprecedented reconfiguration capabilities (center frequency, bandwidth, filter type, and rejection) that is inconceivable with electronic ICs alone \cite{urick2015fundamentals, zhang2015silicon}. However, for their wider adoption in frequency-agile RF photonic receivers or Software Defined Radios (\textbf{SDR}), filter reconfiguration has to be rapid and in-situ, \textit{i.e.}, automatic and free from any manual intervention \cite{choo2018automatic}.

Traditionally, RF photonic systems have been implemented using discrete and bulky photonic components that are power inefficient, expensive and most importantly, are not amenable to integration of complex systems. Silicon Photonics, on the other hand, leverages the unique capabilities of photonic signal processing while takes full advantage of the mature CMOS-like fabrication processes developed by the semiconductor industry \cite{won2010integrating, bogaerts2018silicon, choo2018automatic}. This means, an optical filter realized in a silicon photonic platform is area and power efficient, robust, scalable and can be manufactured on large-scale at a lower cost. However, due to process-induced random variations, any filter designed in a silicon photonic process will deviate from its intended characteristics. Therefore, a scheme is required that not only reconfigures the filter automatically but also is robust against process, voltage and temperature (\textbf{PVT}) variations.

\begin{figure*}[h!]
\centering
\includegraphics[width=0.6\linewidth]{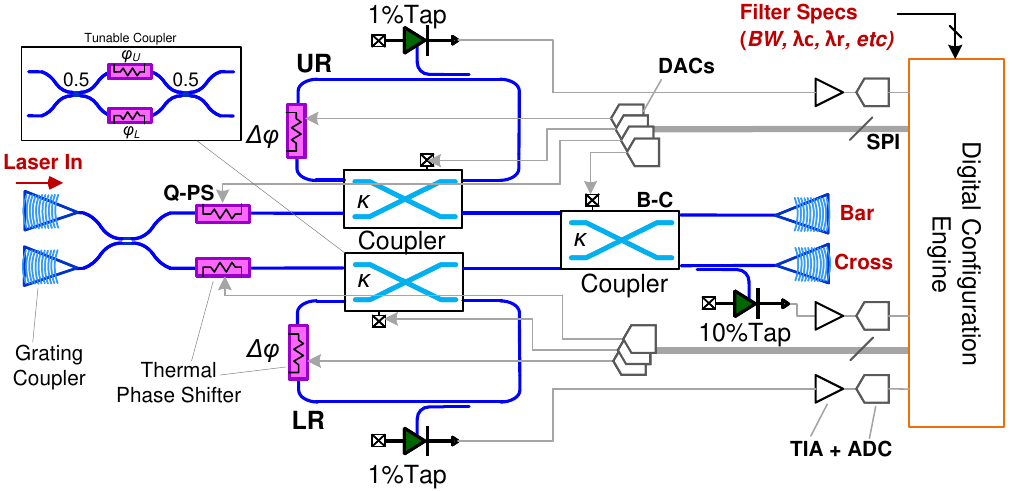}
\caption{Schematic of the silicon photonic filter and control electronics. The outer MZI is loaded with two rings coupled using a tunable coupler and includes microheater phase shifters. Another coupler is used to combine the two arms to compensate for mismatch. Detector taps are placed at the drop port of the rings and on the filter cross port to assist the automatic reconfiguration algorithm. The inset shows the $2\times2$ switch used for realizing tunable couplers. Control electronics (DACs, TIAs and ADCs) are interfaced with a digital configuration engine.}
\label{fig:filter_schematics}
\end{figure*}

Reconfiguration and tuning of integrated optical filters has been pursued in the literature. In ref. \cite{mak2015automatic}, only center frequency tuning was addressed whereas in refs. \cite{mak2016programmable, guan2013cmos, rasras2007demonstration, liao2014integrated}, both center frequency and bandwidth were tunable. However, none of these works pursued out-of-band rejection tuning in the filter. Moreover, the filter reconfiguration was not fully automatic and heavily relied upon manual tuning. The first fully automatic tuning of silicon photonic filter was recently demonstrated in a path-breaking work by Choo et. al. in ref. \cite{choo2018automatic}. However, the reconfiguration process involved the use of bulky and cost prohibitive equipment, \textit{i.e.},  an Optical Vector Network Analyzer (\textbf{OVNA}) to extract ring losses using Jones Matrix based method \cite{choo2018automatic, kim2013linear}.

The fundamental novelty of this work is that it achieves a truly in-situ automatic optical filter reconfiguration solution that has been experimentally demonstrated using a CMOS-compatible silicon-on-insulator (\textbf{SOI}) photonic process. We present an algorithm different from the one in ref. \cite{choo2018automatic} which precludes the use of OVNA and eliminates several time consuming steps during the reconfiguration process. In addition, our filter also uses less number of thermal tuning elements than \cite{choo2018automatic}. The designed filter has a compact form factor and is the first such filter fabricated in AIM Photonics' Active Photonic process. Furthermore, a simplified analytical framework for the design of reconfigurable filters using analog switch-based couplers is provided and the filter PIC simulation leverages our open-source simulation code \cite{saxenaPICTorch22}.
 
\section{Filter design, fabrication and Packaging}

\subsection{Filter Topology}
 The schematic of the second-order filter topology used in this work is shown in \textbf{Fig. \ref{fig:filter_schematics}}. The input continuous-wave (\textbf{CW}) laser is fed to both arms of an outer Mach Zehnder Interferometer (\textbf{MZI}) through a 3dB coupler. Each arm of the MZI is loaded with a microring (UR/LR) via a tunable coupler (UR-C/LR-C) and quadrature phase-shifter (Q-PS). The two tunable couplers are in turn realized using a $2\times2$ MZI switch as shown in the inset of \textbf{Fig. \ref{fig:filter_schematics}}. Another tunable coupler (Back coupler, B-C) is used at the end of the MZI to allow control over the residual imbalance in the optical field between the two arms of MZI caused by process variations. To monitor the ring resonance, a 1\% tap followed by an on-chip Ge-photodetector (\textbf{PD}) is used in each ring. Although this directly translates to increased passband loss of the filter, these taps are an integral part of the automatic software reconfiguration algorithm \cite{choo2018automatic}. Another 10\% monitor tap is used at the cross port of the filter for out-of-band rejection tuning. The automatic calibration and tuning algorithm are implemented with the help of on-chip \textit{thermo-optic phase-shifters} or \textit{microheaters}. Both arms of the outer MZI, both the rings, and all three tunable couplers employ microheaters.

As mentioned earlier, the three tunable couplers used in the RF photonic filter are realized using $2\times2$ analog MZI switches. The coupling matrix, $C$, for these switches is expressed as \cite{capmany2020programmable} 

\begin{align}
C=
je^{j\phi_A} \begin{bmatrix} 
\sin(\phi_D) & \cos(\phi_D)  \\ 
\cos(\phi_D)  & -\sin(\phi_D) 
\end{bmatrix}\\
\triangleq
je^{j (\cos^{-1}(k)+\phi_{A_0})} 
\begin{bmatrix} 
\sqrt{t} & \sqrt{k}  \\ 
\sqrt{k}  & -\sqrt{t} 
\end{bmatrix} \label{eqn:BTU1}
\end{align}

Here, $\phi_A=\frac{\phi_U+\phi_L}{2} + \phi_{A_0}$ is the common-mode phase shift,  $2\phi_D=\phi_U-\phi_L$ is the differential phase shift, and $k$ and $t$ are the cross and through optical power coupling coefficients respectively. The additional phase $\phi_{A_0} = \frac{2\pi c \Delta \tau}{n_{\text{eff}}\lambda}$ is due to the propagation delay, $\Delta\tau$, in the switch due to its physical length. Here, $n_{\text{eff}}$ is the effective index. Moreover, $\phi_U$ and $\phi_L$, are the phase shifts induced in the upper and lower arm microheaters in the switch respectively.  By tuning $\phi_D$, we obtain the power cross-coupling coefficient 

\begin{align} 
    k &= cos^2(\phi_D) = \frac{1}{2}[1+\cos(2\phi_D)] \label{eqn:coupler_sine} \\
      &\triangleq \frac{1}{2}[1+\cos(\varphi_0 + \gamma P_c)] \label{eqn:coupler_Pc}
\end{align}

and the through-coupling coefficient, $t=\sin^2(2\phi_D)$. The (cross-)coupling coefficient $k$ is tuned by applying microheater electrical power, $P_c$, to one of the two microheaters in the switch (i.e. $\phi_U=2\phi_D$ and $\phi_L=0$). This implies the common-mode phase shift is given by

\begin{equation}  \label{eqn:phi_A1} 
   \phi_A=\frac{\phi_U}{2} + \phi_{A_0}= \cos^{-1}(k)+\phi_{A_0}
\end{equation}

Also, in \textbf{Eq. \ref{eqn:coupler_Pc}}, $\varphi_0$ represents the random phase offset in the switch and $\gamma$ is a proportionality constant relating the applied power to the thermo-optic phase-shifter.

The z-domain through (all-pass) and drop (bandpass) transfer functions of a single ring in the filter seen in \textbf{Fig. \ref{fig:filter_schematics}} are derived as-

\begin{equation}  \label{eqn:H_thru} 
    A_{thru}(z) = e^{j\phi_A} \cdot \frac{j\sqrt{t}-\sqrt{t_m}a e^{j(\phi + \phi_A)} z^{-1}}{1+j \sqrt{t t_m} a e^{j(\phi+\phi_A)}z^{-1}}
\end{equation}

\begin{equation}  \label{eqn:H_drop} 
    A_{drop}(z) = \frac{\sqrt{k k_m} e^{j\phi_A} z^{-1/2}}{1+j \sqrt{t t_m} a e^{j(\phi+\phi_A)}z^{-1}}
\end{equation}






where $k_m=0.01$ and $t_m=0.99$ are the cross- and through-coupling coefficients for the 1\% monitor, and $a=e^{-\alpha L}<1$ is the loss factor in the ring and $L$ is the ring length. Here, $\alpha = 10^{2-\frac{\alpha_{w}}{20}}$, where the single-mode SOI waveguide loss $\alpha_{w}$ is 2-2.5dB/cm  \cite{APSUNY_library}.  

\begin{figure*}[!ht]
\centering
\includegraphics[width=0.95\linewidth]{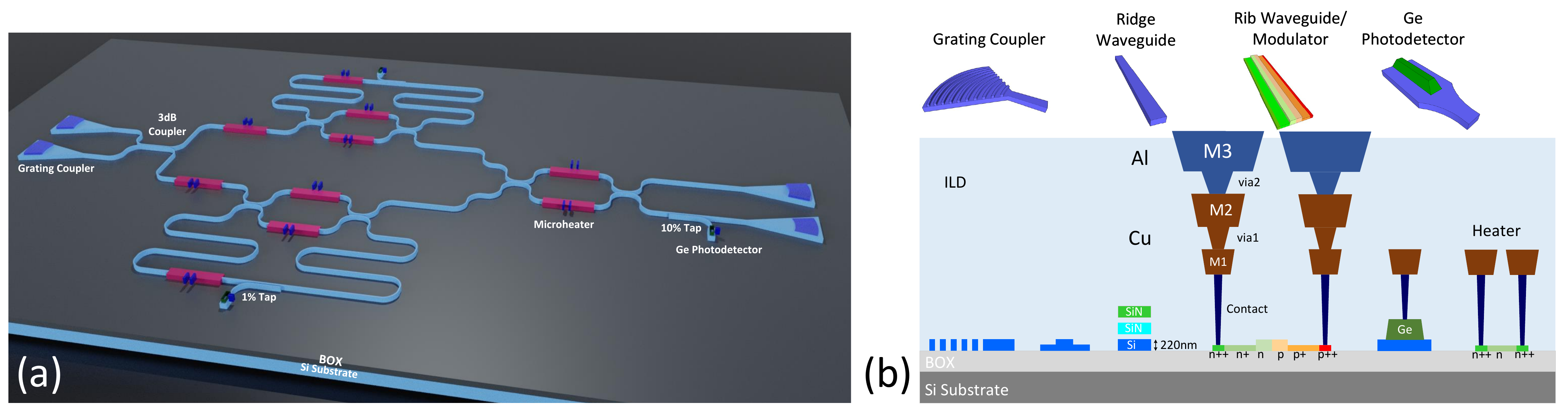}
\caption{(a) A 3D visualization of the filter fabricated in an Active SiP process, (b) cross-section of the SiP process featuring SOI and nitride waveguides, Ge detectors and three metal layers for routing.}
\label{fig:process3D}
\end{figure*}

 A desired filter polynomial, G(z), can be synthesized using  the coupled all-pass decomposition (\textbf{APD}) method by employing sum and difference of two all-pass filters (\textbf{APF}),  $A_1$ and $A_2$, as  $G(z) = \frac{1}{2} (A_1(z) + A_2(z))$ and $H(z) = \frac{1}{2} (A_1(z) - A_2(z))$. For even-order filters, we have $A_{2}(z)=A_{1}^*(z^*)$ \cite{regalia1988digital,madsen1998efficient}. 
 The $2^{nd}$-order APD filter seen in \textbf{Fig. \ref{fig:filter_schematics}} had two ring APF arms and the sum or difference is realized using the back-coupler, B-C. The resulting filter output bar and cross transfer functions are 
 
 \begin{align}  
    G(z),H(z) &= \frac{j}{\sqrt{2}} 
                \Big( \sqrt{k_{bc}}e^{j\phi_{q,ps}} A_{thru}(z)\Big|_{k_1, \phi_1} \nonumber \\
                & \pm \sqrt{t_{bc}}e^{-j\phi_{q,ps}} A_{thru}(z)\Big|_{k_2, \phi_2} \Big)  \label{eqn:H_filter2}
\end{align}
 
 In our notation, $k_n$ and $\phi_n$ are the power coupling coefficient and phase shift in the $n^{th}$ ring, $k_{bc}$ and $t_{bc}=1-k_{bc}$ are the back-coupler cross and through coupling coefficients, and $\phi_{q,ps}$ is the quadrature phase shift.

\subsection{Filter Synthesis} 
Filter design starts with the synthesis of a polynomial, G(z), for the specified filter type and specifications. Then the denominators of the coupled all-pass polynomials, $D_{1,2}(z)$, corresponding to G(z), and constant phase $\beta$ were obtained using the \textbf{\textit{tf2ca}} function in Matlab\texttrademark. Then numerators were obtained using $N_1(z)=\textbf{\textit{fliplr}}(D_1^*(z))$ and $N_2(z)=\textbf{\textit{fliplr}}(D_2^*(z))$. This yielded the two all-pass transfer functions $A_1(z)=e^{j\beta}\frac{N_1(z)}{D_1(z)}$ and $A_2(z)=A_1^*(z^*)=e^{-j\beta}\frac{N_2(z)}{D_2(z)}$.  
Next, the all-pass transfer functions $A_{1,2}(z)$ are mapped to the (cascade of) ring resonators in the upper and lower arms. If the roots of denominators $D_{1}(z)$, i.e. the poles of $A_{1}(z)$, are $p_n$, then the cross-coupling coefficients are determined using $k_n=(1-|p_n|^2)$ and $t_n=|p_n|^2$. The phase $\phi_n=\angle p_n$. The quadrature phase shift for the upper arm is obtained as $\phi_{q,ps}=\beta-\sum_n \phi_n$. The lower arm coupling coefficients are identical to those of the upper arm, and $\phi_n$ and $\phi_{q,ps}$ are of opposite sign, representing the conjugate APF responses. In this work, we used a $2^{nd}$-order Butterworth response with a normalized cutoff frequency of $\omega_n=0.13$. The synthesized filter coefficients for the filter specifications are listed in  Table \ref{tab:filter_coefficients}. 

\begin{table}[htbp]
\centering
\caption{Filter coefficients for the second-order APD-type optical filter.}
\begin{tabular}{cccc}
\hline
Ring (n) & $k_{n}$ & $\phi_{n}$ & $\phi_{q,ps}$ \\
\hline
1 (UR) & 0.4385 & -0.2969 & Top: -1.6137\\
2 (LR) & 0.4385 & 0.2969 & Bot: 1.6137\\
\hline
\end{tabular}
  \label{tab:filter_coefficients}
\end{table}

\subsection{PIC Design and Fabrication} 

The filter was designed and fabricated in AIM Photonics foundry's 300mm Active Photonics process \cite{aimphot}. This design was a part of AIM's  MPW run in June, 2019. \textbf{Fig. \ref{fig:process3D}} shows a 3D visualization of the filter schematic seen in \textbf{Fig. \ref{fig:filter_schematics}}, and the process cross-section. This process features silicon-on-insulator (SOI) rib and ridge waveguides, silicon nitride waveguides, escalators, modulator doping, Germanium (Ge) detectors, and three metal layers for routing \cite{aimphot, timurdogan2018aim, fahrenkopf2019aim}.  The PIC layout employed several process-optimized components from the PDK library \cite{APSUNY_library, timurdogan2018aim, fahrenkopf2019aim}. Grating couplers (\textbf{GC}) were used for vertical coupling of light into the PIC. Single-mode silicon waveguides of 480nm width and 220nm height were used for routing and interconnecting the optical components in TE polarization. The analog  $2\times2$ switches were realized using doped silicon waveguide microheater sections and 3-dB couplers \cite{watts2013adiabatic}, and with a measured thermo-optic time-constant of ~$\sim 15\mu$s.
The physical lengths of the microheaters and $2\times2$ switches are around 100$\mu m$ and 550$\mu m$, respectively.

\begin{figure}[ht!]
\centering 
(a)
\includegraphics[width=0.9\linewidth]{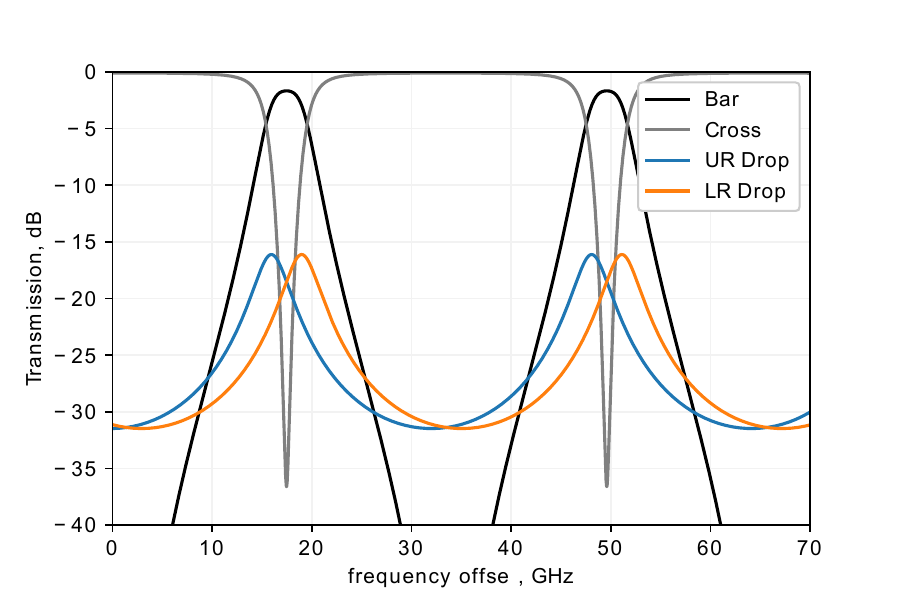}
(b)
\includegraphics[width=0.92\linewidth]{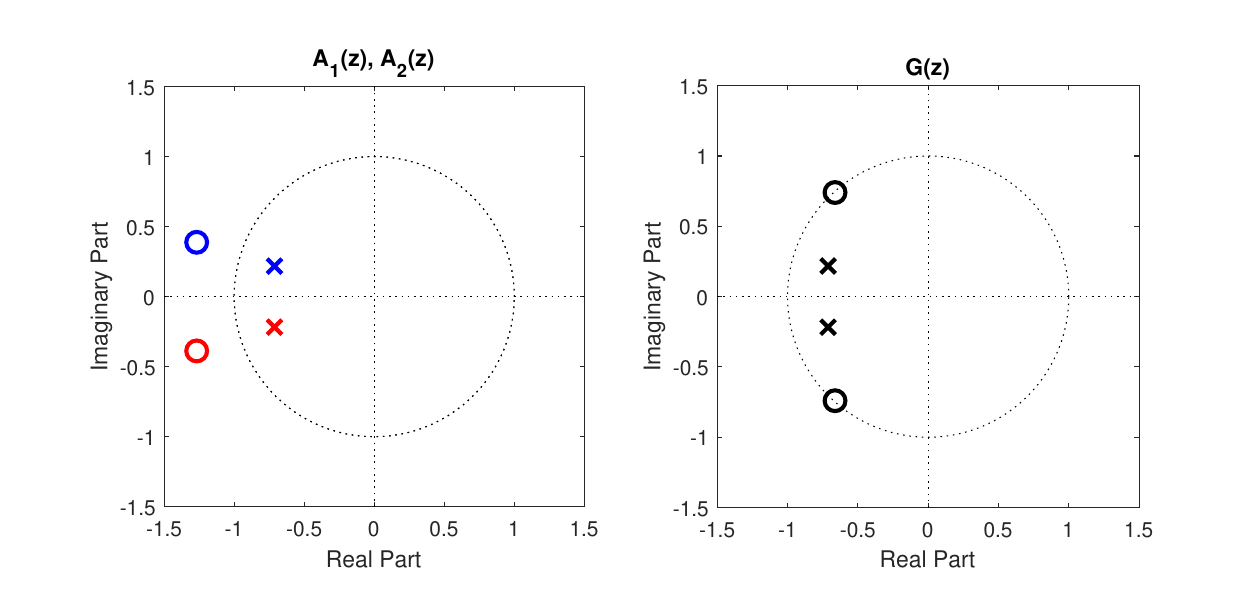}
\caption{ (a) Transmission response of the second-order APD-type filter simulated using custom PhotonTorch-based code and the SiP process data. (b) Complex z-plane pole-zero plots for $G(z)$ and $A_{1,2}(z)$. }
\label{fig:filter_sim1}
\end{figure}

The filter was simulated using Lumerical Interconnect \cite{lumericalInterconnect} with the AIM PDK library, as well as our open-source Python-based PIC simulator  \cite{saxenaPICTorch22} built using the PhotonTorch photonic simulation framework \cite{laporte2019highly}. \textbf{Fig. \ref{fig:filter_sim1}}(a) shows the simulated bar and cross responses of the filter after including all PIC parameters including the losses and delays. The peak of the optical monitors are aligned to the desired resonance frequency (or wavelength), corresponding to the ring phase, $\phi_n$, for each of the rings. The simulated passband loss using the synthesized filter coefficients was $\sim$2.2 dB due to the ring losses with a 3dB bandwidth (BW) of 3 GHz and an out-of-band rejection of $>$30dB. The pole-zero plots for APFs, and bar and cross responses, i.e. $G(z)$ and $H(z)$ respectively, are shown in \textbf{Fig. \ref{fig:filter_sim1}}(b).

The filter layout was compacted as much as possible to share the photonic chip with other RF photonic circuits. The rings were routed in a compact serpentine fashion with a 30$\mu$m bend radius for lower loss and reflections. The resulting total physical length of the rings were $\sim$2200$\mu m$ each, with a corresponding free-spectral range (\textbf{FSR}) of around 31GHz. The monitor taps were realized using 1\% and 10\% couplers, Ge waveguide detectors and waveguide terminations.  The filter die micrograph is shown in \textbf{Fig. \ref{fig:filter_packaging}(a)}. The filter core layout occupied 1.12 mm$^2$ area on this chip.

\subsection{Packaging and Electronic Back-end} 
\begin{figure}[ht!]
\centering
\includegraphics[width=1\linewidth]{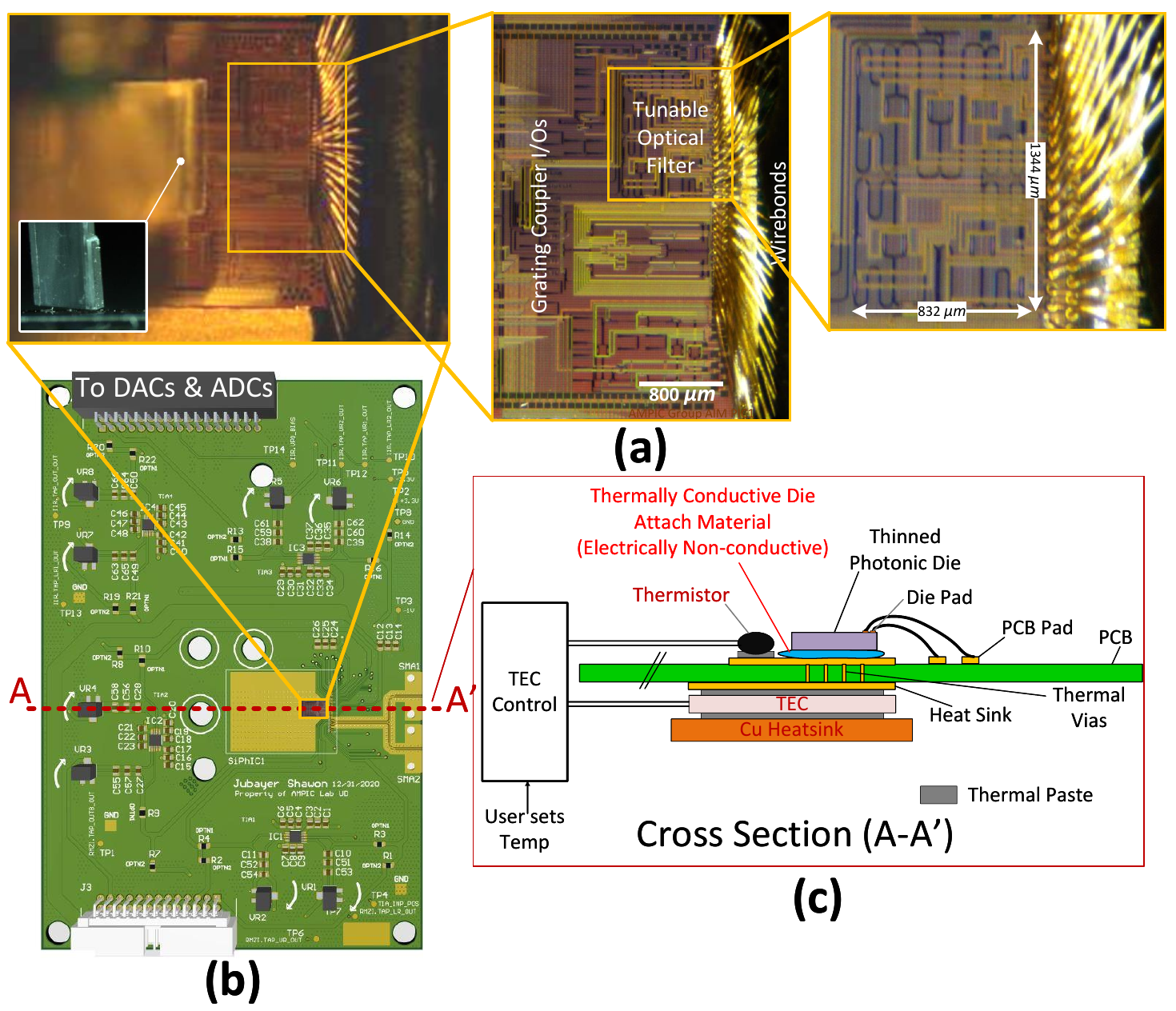}
\caption{(a) Chip micrograph of the silicon photonic filter fabricated in the AIM Photonics process. (b) The chip is attached and wire-bonded onto a custom designed printed circuit board (PCB) using Chip-On-Board (COB) packaging.  The PCB contains transimpedance amplifiers (TIAs) and ribbon cable interface to another electronic board with DACs and ADCs.  (c) The COB assembly is co-packaged with a thermistor on the top-side and a Peltier cell on the back side of the PCB with an external TEC controller board in a closed loop.}
\label{fig:filter_packaging}
\end{figure}

The fabricated PIC die was polished down to 150$\mu m$ and packaged in a chip-on-bard (\textbf{COB}) assembly as shown in \textbf{Fig. \ref{fig:filter_packaging}(b \& c)}. A Peltier cell and thermistor were used along with a thermo-electric cooler (\textbf{TEC}) controller in a closed-loop to stabilize the temperature of the chip and to minimize the thermal crosstalk among the on-chip tuning elements. The combination of die thinning and TEC at the bottom of the die provides effective thermal isolation by creating a prominent thermal gradient in the vertical direction \cite{choo2018automatic}. 

The electrical pads were placed on two rows on the East edge of the PIC which were wire-bonded with two rows of PCB pads. The optical monitors were connected to the on-board transimpedance amplifiers (\textbf{TIAs}), whose outputs were interfaced with commercial off the shelf (COTS) 16-bit analog-to-digital converters (\textbf{ADCs}) using a ribbon cable. These pads provide 16-bit digital-to-analog converters (\textbf{DACs}) interfaces to the microheaters. All the DACs and ADCs communicate via SPI interface with a microcontroller, which provides a software abstraction to the algorithm code.

The operating voltage range of the microheaters is from 0 to 5V and requires a power of $P_\pi\approx 30$mW for an optical phase shift of $\pi$. These voltage ranges are CMOS-compatible and the microheater driver circuits can be implemented using stacked high-voltage I/O transistors with a 5V supply voltage in 65nm CMOS (or similar) process technology \cite{li20183d,zhu2015design}. The TIA interface with the PDs and the 16-bit DACs and ADCs can be designed using the standard transistors in CMOS, allowing future integration of all electronics on the same chip.

\section{In-Situ Component Parameter Extraction}

To reconfigure the SiP filter, device parameters such as the coupling ratio of the tunable couplers ($k_{n}$) and the ring phase bias ($\phi_{n}$) for the selected center wavelength ($\lambda_c$) need to be configured. Here, $n=1,2,\ldots N$ stands for the ring number. These parameters are controlled by the DAC voltage (or power). Due to PVT variations in the photonic chip, there is a significant variation in the response of the tunable couplers and the phase shifters with respect to the applied power. Thus, no two couplers (or phase shifters) produce the same coupling (phase shift) for a given applied DAC voltage. 

The phase shift, $\phi_n$, produced by the microheaters linearly depends upon the applied electrical power, $P_n$, as 

\begin{equation}  \label{eqn:P_n1} 
   \phi_n=\gamma P_n =  \frac{\pi P_n}{P_{\pi}}
\end{equation}

However, the microheater resistance, $R$, exhibits nonlinearity due to self-heating. Thus, applied electrical power has a nontrivial nonlinear dependence on the applied DAC voltage, $P_n=f(v_n) \ne \frac{v_n^2}{R}$. This $P_n$ \textit{vs} $v_n$ characteristics must also be extracted so that we can accurately configure component parameters on the chip using the DACs. This necessitates an one-time in-situ pre-characterization of each of these components, and subsequently the configuration of them, as they are used. In this section, we provide algorithms for these pre-characterization routines and configuration.

Each of these component characterization have two steps- Part I and II. The first part, Part I, for either $k_{n}$ or $\phi_{n}$ extraction represents the component pre-characterization phase. On the other hand, during the automatic reconfiguration phase, this pre-characterized data is used for in-situ estimation and configuration of $k_{n}$ and $\phi_{n}$ parameters in the on-chip filter. Part II algorithms for both $k_{n}$ and $\phi_{n}$ represent these in-situ estimation and configuration.

It is important to note that Part I is a one-time operation and does not need to be repeated again as long as the filter ambient temperature is kept the same as the filter pre-characterization temperature. This is ensured by using the TEC setup seen in \textbf{ Fig. \ref{fig:filter_packaging}}. On the other hand, Part II algorithms are invoked each time the filter is reconfigured or tuned to a different center frequency (or wavelength).

\subsection{Coupling Ratio ($k_{n}$) Extraction} 

\begin{figure}[hbt!]
\footnotesize
\centering
\includegraphics[width=1\linewidth]{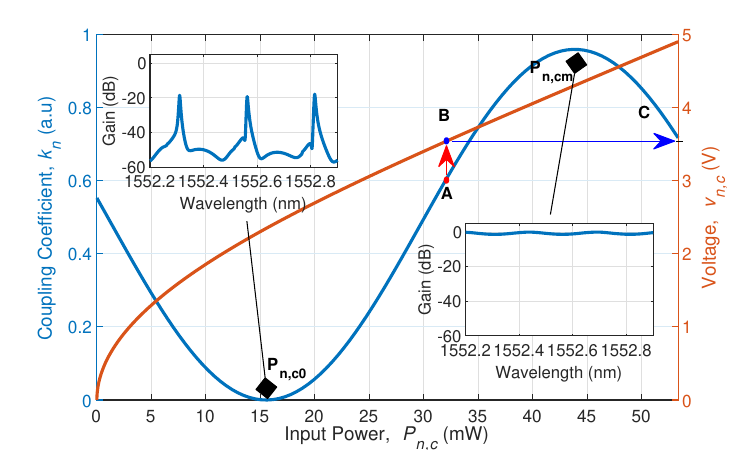}
\caption{(Illustration of Algorithms \ref{alg:ccoeffp1} and \ref{alg:ccoeffp2}): Coupling coefficient, $k_{n}$, of a tunable coupler with a process induced random phase offset and the microheater voltage as a function of input power, $P_{n,c}$, as expressed by \textbf{Eq. \ref{eqn:coupler_Pc}}. The top-left and bottom-right insets show normalized wavelength response of the ring drop port with its coupler set at zero ($P_{n,c}=P_{n,c0}$ \& $k=0$) and maximum coupling ratio ($P_{n,c}=P_{n,cm}$ \& $k=k_{n,max}$), respectively. Any desired value for $k=k_n$ (labeled as \textbf{A}) can be in-situ configured by applying the corresponding input power, $P_{n,c}^*$ (label \textbf{B}), which in turn is translated to the DAC voltage, $v_{n,c}^*$ (label \textbf{C}).}
\label{fig:zero_max_coupling}
\end{figure}

During the initial pre-characterization routine, \textbf{Algorithm \ref{alg:ccoeffp1}} is executed for each of the $n$ rings in the filter. Here, to be able to configure the desired coupling ratio ($k_{n}$) of the tunable coupler of the n$^{th}$ ring, the wavelength response of the ring drop-port is recorded (using the monitor PDs) for a range of coupler input voltages ($v_{n,c}$). From this data, the voltage ($v_{n,cm}$) corresponding to the \textbf{maximum coupling ratio ($k_{n,max}$)} is recorded. The maximum coupling is identified by observing the quality factor (Q) of the ring drop-port resonance. Optical cavity dynamics dictate that as coupling ratio approaches unity, ring Q decreases \cite{bogaerts2012silicon}. This provides the basis for identifying the $k_{n,max}$ by observing the \textbf{minimum Q} (as shown in the bottom-right inset of \textbf{Fig. \ref{fig:zero_max_coupling}}). On the other hand, zero-coupling voltage ($v_{n,c0}$) is identified simply by finding the $v_{n,c}$ corresponding to lowest monitor output at a fixed $\lambda \in \Lambda$ (any wavelength within one FSR in the wavelength range of interest, \textit{i.e.} $\Lambda$). This is possible due the to very low optical power coming through the drop-port when zero-coupling is in place (as shown in the top-left inset of \textbf{Fig. \ref{fig:zero_max_coupling}}). Afterwards, the power ($P_{n,c}$) delivered to n$^{th}$ coupler microheater at different $v_{n,c}$ is recorded using a Keysight B2902A source measurement unit (\textbf{SMU}), and thus the $P_{n,c}$ \textit{vs} $v_{n,c}$ characteristics are obtained. The SMU can also be realized using on-board electronics for portability \cite{EasySMU2021}.

Utilizing this data rather than simply relying on the quadratic relationship between voltage and power \cite{li202012} ensures that the algorithm is insensitive to the electrical non-linearity of on-chip microheaters. Subsequently, the zero-coupling power ($P_{n,c0}$) and max-coupling heater power ($P_{n,cm}$) are obtained, and thus the power ($P_{n,{c_{\pi}}}$) required to obtain a $\pi$ phase shift in the coupler microheaters is determined as $P_{n,{c_{\pi}}} = P_{n,cm} - P_{n,c0}$. This $P_{n,{c_{\pi}}}$ will subsequently serve as a reference value for the tuning algorithm to operate at different wavelengths within the FSR of the rings. It is important to note that by utilizing zero and maximum coupling rather than the critical coupling as in \cite{choo2018automatic}, we avoid repeated re-centering of the ring resonance at the filter center frequency ($\lambda_c$) during the $k_{n}$ extraction step. Also as mentioned earlier, use of an OVNA is avoided by precluding the ring loss measurements at critical coupling for each of the rings as in the prior art in \cite{choo2018automatic}. This considerably speeds up the $k_{n}$ extraction step in the filter reconfiguration algorithm ($\sim$2X reduction in time) and enables in-situ tuning that can be implemented using PCB-level or chip-scale electronics.

\begin{algorithm}
\footnotesize
\caption{Coupling ratio extraction (Part I)} \label{alg:ccoeffp1}
\begin{algorithmic}[1]
\State \textbf{Start}
\State \textbf{Get} n$^{th}$ Ring drop-port response, $f(\lambda,v_{n,c})$ \& estimate $Q$
\State $v_{n,cm}$, $P_{n,cm}\gets$ \textbf{Find} $v_{n,c}$, $P_{n,c}$ corresponding to \textbf{min}(Q)
\State \textbf{Set} laser wavelength at $\lambda \in \Lambda$
\State \textbf{Get} n$^{th}$ Ring drop-port response, $f(v_{n,c})$
\State $v_{n,c0}$, $P_{n,c0}\gets$ \textbf{Find} $v_{n,c}$, $P_{n,c}$ for zero coupling from \textbf{min}$(f(v_{n,c}))$
\State $P_{n,{c_{\pi}}} \gets P_{n,cm} - P_{n,c0}$
\State \textbf{Save} $P_{n,{c_{\pi}}}$
\State \textbf{Save} $v_{n,c}$ \textit{vs} $P_{n,c}$ characteristics
\State \textbf{End}
\end{algorithmic}
\end{algorithm}

During the in-situ reconfiguration routine in \textbf{Algorithm \ref{alg:ccoeffp2}}, first the $P_{n,c0}$ is recorded, and depending on the $\lambda_c$, this may not be the same as the $P_{n,c0}$ seen in \textbf{Algorithm \ref{alg:ccoeffp1}}. Then, the raised-cosine coupling ratio \textit{vs} power curve of the tunable coupler \cite{choo2018automatic}  from \textbf{Eq. \ref{eqn:coupler_Pc}} and illustrated in \textbf{Fig. \ref{fig:zero_max_coupling}} is fitted to the ($P_{n,c0}$, 0) \& ($P_{n,c0} + P_{n,{c_{\pi}}}$, $k_{n,max}$) data points. From this fitted curve, power ($P_{n,c}^*$) required to set any desired coupling ratio ($k_{n}$) is estimated  as shown in \textbf{Fig. \ref{fig:zero_max_coupling}}. Afterwards, from the pre-characterization $v_{n,c}$ \textit{vs} $P_{n,c}$ data, the DAC voltage ($v_{n,c}^*$) needed to set a desired $k_{n}$ is determined. Here, $k_{n,max}$ is the maximum coupling coefficient and due to the intrinsic loss in the coupler, it never reaches unity. For our tunable coupler, it is found to be 0.959 from the foundry PDK. Furthermore, for the purpose of the filter and its usable frequency range for a free-spectral range (FSR) of 31 GHz (or 0.249 nm), the couplers can be considered broadband. Thus, all the extracted parameters are usable at any frequency (or wavelength) within the FSR.

\begin{algorithm}
\footnotesize
\caption{Coupling ratio extraction (Part II)}\label{alg:ccoeffp2}
\begin{algorithmic}[1]
\State \textbf{Input}: $k_{n}\gets$ User specifies the desired coupling coefficient
\State \textbf{procedure} \textbf{ccoeff}(n, $\lambda_c$, $k_{n})$
\State \qquad\textbf{Set} laser wavelength at $\lambda_c$
\State \qquad\textbf{Get} n$^{th}$ Ring drop-port response, $f(v_{n,c})$
\State \qquad$v_{n,c0}$, $P_{n,c0}\gets$ \textbf{Find} $v_{n,c}$, $P_{n,c}$ for zero coupling from \textbf{min}$(f(v_{n,c}))$
\State \qquad\textbf{Fit} $k_n$ \textit{vs} $P_{n,c}$ in \textbf{Eq. \ref{eqn:coupler_Pc}} using ($P_{n,c0}$, 0) \& ($P_{n,c0} + P_{n,{c_{\pi}}}$, $k_{n,max}$) data points
\State \qquad$P_{n,c}^*\gets$ \textbf{Find} power for $k_{n}$ from the above curve-fit
\State \qquad$v_{n,c}^*\gets$ \textbf{Find} voltage corresponding to $P_{n,c}^*$ from $v_{n,c}$ \textit{vs} $P_{n,c}$ data
\State \qquad\textbf{Return} n$^{th}$ coupler DC bias, $v_{n,c}^*$
\State \textbf{End procedure}
\end{algorithmic}
\end{algorithm}

\subsection{Ring Phase ($\phi_n$) Extraction}

Now we describe the one-time pre-characterization routine (i.e. \textbf{Algorithm \ref{alg:ringbiasp1}}) for the ring phase bias. First, all the rings except for the n$^{th}$ ring are detached from the filter by setting their corresponding couplers to zero-coupling. 
This ensures that the cross-port of the filter only shows the resonance response of the n$^{th}$ ring. Then, the voltage ($v_{n,r_{\pi}}$) required for a $\pi$ phase shift in the ring microheater was extracted by monitoring the cross port of the filter and shifting the resonance by full FSR (corresponding to $2\pi$ phase-shift in the ring) for the applied voltage. Afterwards, the power ($P_{n,r}$) delivered to n$^{th}$ ring heater at different $v_{n,r}$ is recorded using the SMU and $P_{n,{r_{\pi}}}$ is obtained. As mentioned before, utilizing $v_{n,r}$ \textit{vs} $P_{n,r}$ data ensures that the algorithm is insensitive to the process-dependent electrical nonlinearity of the on-chip microheaters.

\begin{algorithm}
\footnotesize
\caption{Ring phase extraction (Part I)}\label{alg:ringbiasp1}
\begin{algorithmic}[1]
\State \textbf{Start}
\State \textbf{Set} $v_{i,c}\gets$ \textbf{Find} $v_{i,c0}$ for all couplers except when $i=n$
\State \textbf{Get} filter bar-port response, $f(\lambda,v_{n,r})$
\State $v_{n,r_{\pi}}$, $P_{n,r_{\pi}}\gets$ \textbf{Find} $v_{n,r}$, $P_{n,r}$ for $\pi$ phase-shift, obtained using a full-FSR sweep
\State \textbf{Save} $P_{n,{r_{\pi}}}$
\State \textbf{Save} $v_{n,r}$ \textit{vs} $P_{n,r}$ data
\State \textbf{End}
\end{algorithmic}
\end{algorithm}

During the automatic calibration routine in \textbf{Algorithm \ref{alg:ringbiasp2}}, pre-characterized $P_{n,{r_{\pi}}}$, and  $v_{n,r}$ \textit{vs} $P_{n,r}$ characteristics of the ring microheater is utilized. Then, from the phase-shift desired by the user, $\phi_{n}$, the required microheater power change, $\Delta P_{n,r}$, is estimated using \textbf{Eq. \ref{eqn:P_n1}} as

\begin{equation}  \label{eqn:delta_Pnr} 
    \Delta P_{n,r}=\frac{\phi_{n}P_{n,{r_{\pi}}}}{\pi}
\end{equation}

Meanwhile, the voltage ($v_{n,res}$) for which the n$^{th}$ ring resonance is aligned with $\lambda_c$ is obtained by sweeping the ring microheater voltage ($v_{n,r}$) and finding the maximum drop port response. From this data, the power ($P_{n,res}$) at which ring resonance is aligned with $\lambda_c$ is obtained. Then, the power ($P_{n,r}^*$) required to set the ring at desired $\phi_{n}$ is estimated as $P_{n,r}^* = P_{n,res} + \Delta P_{n,r}$ and the corresponding ring microheater voltage ($v_{n,r}^*$) is applied.

\begin{algorithm}
\footnotesize
\caption{Ring phase extraction (Part II)}\label{alg:ringbiasp2}
\begin{algorithmic}[1]
\State \textbf{Input}: $\phi_{n}\gets$ User specifies the desired ring phase
\State \textbf{procedure} \textbf{ringbias}(n, $\lambda_c$, $\phi_{n}$)
\State \qquad$\Delta P_{n,r}\gets \frac{\phi_{n}P_{n,{r_{\pi}}}}{\pi}$
\State \qquad\textbf{Set} laser wavelength at $\lambda_c$
\State \qquad\textbf{Get} n$^{th}$ Ring drop-port response, $f(v_{n,r})$
\State \qquad$v_{n,res}$, $P_{n,res}$$ \gets$ \textbf{Find} $v_{n,r}$, $P_{n,r}$ for \textbf{max}$(f(v_{n,r}))$ \Comment{resonance}
\State \qquad$P_{n,r}^* \gets P_{n,res} + \Delta P_{n,r}$
\State \qquad$v_{n,r}^*\gets$ \textbf{Find} voltage for $P_{n,r}^*$ from the $v_{n,r}$ \textit{vs} $P_{n,r}$ data
\State \qquad\textbf{Return} n$^{th}$ ring heater DC bias $v_{n,res}$ \& $v_{n,r}^*$
\State \textbf{End procedure}
\end{algorithmic}
\end{algorithm}

\section{Automatic Tuning of Filters}

The filter tuning algorithm (\textbf{Algorithm \ref{alg:main_algorithm}}) starts off with the user specifying the desired filter specifications. The APD filter synthesis algorithm \cite{madsen1998efficient,shawon2019rapid,shawon2020rapid} translates these user-defined specifications (filter type, order, bandwidth, center frequency, out-of-band suppression) to filter parameters ($k_{n}$, $\phi_{n}$ \& $\phi_{q,ps}$) for all rings and MZI arms. Then the $\textbf{ccoeff}(n, \lambda_c, k_{n})$ routine from \textbf{Algorithm \ref{alg:ccoeffp2}} is invoked and desired coupling coefficients for all ring couplers are configured by setting the microheater voltages of the ring tunable couplers to $v_{n,c}^*$. 
Next, the $\textbf{ringbias}(n, \lambda_c, \phi_{n})$ routine from \textbf{Algorithm \ref{alg:ringbiasp2}} is invoked to set all the ring biases ($\phi_{n}$) with respect to the $\lambda_c$ at their desired values. It's  important to note that due to thermal crosstalk between on-chip microheaters, every time a ring phase ($\phi_n$) is set, previously aligned rings get de-tuned and several iterations are required accurately to set all the $\phi_{n}$ values.

To improve the out-of-band rejection of the filter, we maximize the 10\% monitor tap placed at the cross-port of the filter at $\lambda_r$ by tuning the MZI quadrature-phase-shifter (Q-PS) and the back-coupler (B-C). Here, $\lambda_r$ is the wavelength where a null in the filter response can be enforced.

However, the out-of-band-rejection calibration via Q-PS and B-C microheaters shifts the resonance of the previously tuned rings, thus altering the center wavelength (or frequency) of the filter response. To perform center-wavelength correction, rings, Q-PS and B-C are tuned subsequently and several iterations of tuning ($n$ rings, Q-PS and B-C) are needed to reach the thermal steady-state of the PIC. When the required voltage stimulus is stabilized within a tolerance range (indicating a thermal steady-state), the filter tuning is complete. 

It's important to note that the coupling ratio of the tunable couplers are relatively stable in presence of thermal perturbation. This is due to the fact that thermal perturbation coming from other microheaters affect both microheaters of the tunable couplers almost equally, thus inducing a `common-mode' phase shift that does not alter the coupling ratios which only depends upon the differential phase ($\phi_D$). Therefore, it is not required to include ring coupler tuning inside the iteration loop, which in turn speeds up the reconfiguration process. Another key aspect of this algorithm is that it precludes the use of `outer ring' for out-of-band rejection in \cite{choo2018automatic}, thus further reducing configuration time for `outer ring' resonance alignment and also results in a smaller layout area.

\begin{algorithm}[htbp]
\footnotesize
\caption{N$^{th}$ Order Filter reconfiguration Algorithm}\label{alg:main_algorithm}
\begin{algorithmic}[1]
\State \textbf{Start}
\State \textbf{Get} Filter specifications (Type, Order, BW, Rejection, $\lambda_c$, $\lambda_r$) from the user
\State $k_{n}$, $\phi_{n}$, $\phi_{q,ps}\gets$ APD filter synthesis using the specifications
\State \textbf{Set} $n=1$
\State \textbf{while} $n \leq N$ \textbf{do}
\State \qquad\textbf{Invoke} [$v_{n,c}^*$] = ccoeff(n, $\lambda_c$, $k_{n})$ routine
\State \qquad\textbf{Set} $v_{n,c}^*\gets$ DC bias of n$^{th}$ coupler
\State \qquad\textbf{Set} $n=n+1$
\State \textbf{end while}
\State \textbf{Set} $n=1$
\State \textbf{while} $\text{err}_j \geq \text{tol}_j$ \textbf{do}
\State \qquad\textbf{while} $\text{err}_i\geq \text{tol}_i$ \textbf{do}
\State \qquad\qquad\textbf{while} $n\leq N$ \textbf{do}
\State \qquad\qquad\qquad\textbf{Invoke} [$v_{n,res}$, $v_{n,r}^*$] = ringbias(n, $\lambda_c$, $\phi_{n}$)
\State \qquad\qquad\qquad\textbf{Set} $v_{n,res}\gets$ DC bias for n$^{th}$ ring resonance
\State \qquad\qquad\qquad\textbf{Set} $n=n+1$
\State \qquad\qquad\textbf{end while}
\State \qquad\qquad\textbf{Calculate} $\text{err}_i$
\State \qquad\textbf{end while}
\State \qquad\textbf{Set} $v_{n,r}^*\gets$ DC bias of n$^{th}$ ring
\State \qquad\textbf{Set} Laser wavelength at $\lambda_r$
\State \qquad\textbf{Get} filter cross-port PD response, $f(v_{q,ps})$
\State \qquad$v_{q,psm}\gets$ \textbf{Find} $v_{q,ps}$ for \textbf{max}$f(v_{q,ps})$
\State \qquad\textbf{Set} DC bias of the arm phase shifter at $v_{q,psm}$
\State \qquad\textbf{Get} filter cross-port PD response, $f(v_{b,c})$
\State \qquad$v_{b,cm}\gets$ \textbf{Find} $v_{b,c}$ for \textbf{max}$f(v_{b,c})$
\State \qquad\textbf{Set} DC bias of back-end tunable coupler at $v_{b,cm}$
\State \qquad\textbf{Calculate} $\text{err}_j$
\State \textbf{end while}
\State \textbf{End}
\end{algorithmic}
\end{algorithm}

\section{Experimental Results}

To demonstrate the functionality and efficacy of the proposed reconfiguration algorithm, the $2^{nd}$-order Butterworth filter was experimentally reconfigured using the coefficients from \textbf{Table \ref{tab:filter_coefficients}}. This filter is intended for $\sim$30dB out-of-band rejection, an insertion loss (IL) of 2.2 dB, and the 3dB bandwidth (BW) of 3 GHz at 1550nm wavelength. The 1550nm CW laser and input/outputs were coupled into the PIC using a 4-channel polarization maintaining fiber array. The fiber array was automatically aligned to the on-chip GCs. 

\begin{figure}[htbp]
\centering
\includegraphics[width=1\linewidth]{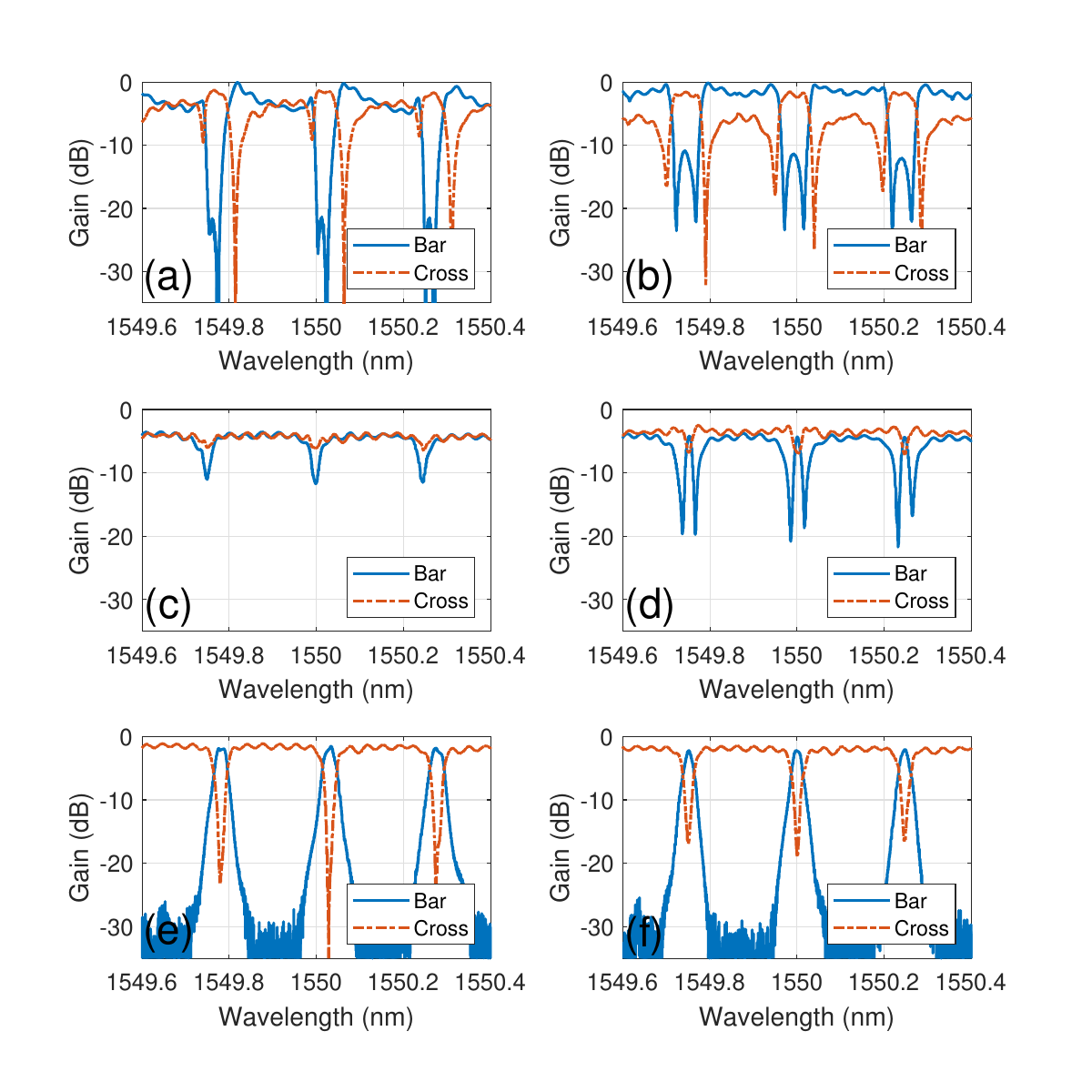}
\caption{Experimentally measured bar and cross port responses (normalized) of the filter: (a) When powered up without any tuning, (b) After all the coupling ratios ($k_n$) of the rings are configured (Algorithm \ref{alg:ccoeffp2}), (c) After aligning the ring resonances to $\lambda_c$ (part of Algorithm \ref{alg:ringbiasp2}), (d) After setting desired phase bias ($\phi_n$) of the rings (part of Algorithm \ref{alg:ringbiasp2}), (e) After Quadrature Phase-Shifter (Q-PS) and Back-Coupler (B-C) tuning, (f) After center frequency correction i.e. several iterations of ring phase bias ($\phi_n$), Q-PS and B-C tuning.}
\label{fig:filter_tuning_steps}
\end{figure}

The experimentally measured response of the filter at different stages of reconfiguration is illustrated in \textbf{Fig. \ref{fig:filter_tuning_steps}}. The observed GC response exhibited $\sim$0.8dB ripple in the passband response, and the passband GC losses were de-embedded from the filter response. Unlike \cite{choo2018automatic}, here in this work, as soon as the Q-PS and/or B-C microheater is tuned, the previously locked UR $\&$ LR rings get de-tuned due to thermal crosstalk and the center frequency of the filter shifts from its intended $\lambda_c$ (in this case, $\lambda_c$ = 1550 nm) as demonstrated in\textbf{ Fig. \ref{fig:filter_tuning_steps}}(d) and (e). Therefore, several iterations are required for automatic tuning of the filter.

\begin{figure*}[htbp]
\centering
\includegraphics[width=0.75\linewidth]{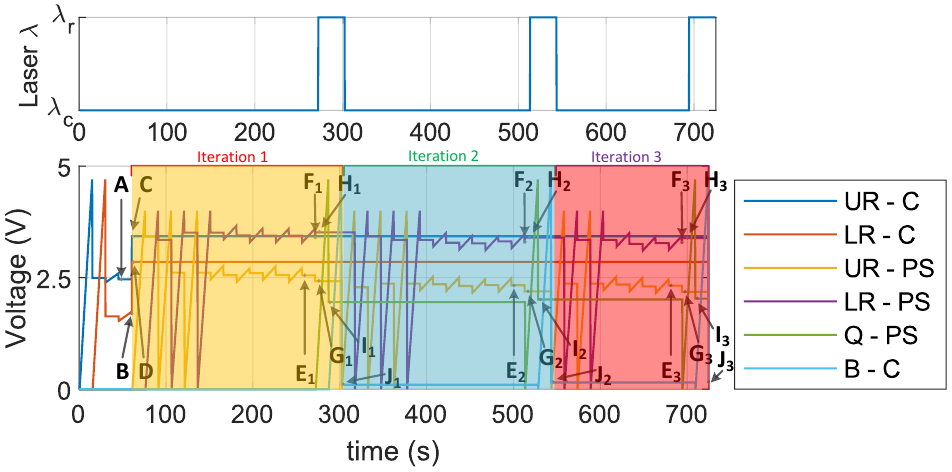}
\caption{DAC sampling and laser wavelength switching at different stages of silicon photonic filter reconfiguration. The corresponding timestamps are explained in detail in Table \ref{tab:dac_timestamps}. Each iteration corresponds to ring phase, quadrature phase-shifter and back-coupler configuration steps. Here, three iterations were needed as the thermal crosstalk from the last back-coupler and quadrature phase-shifter configuration step de-tunes the rings.}
\label{fig:dac_sampling}
\end{figure*}

Since DACs set the microheater voltages (or power), the convergence of the tuning algorithm can be observed in the settling of the DAC output voltages. In \textbf{Fig. \ref{fig:dac_sampling}}, the transient evolution of DAC voltages for all relevant microheaters during the tuning process is presented. \textbf{Fig. \ref{fig:dac_sampling}} also shows the laser wavelength settings ($\lambda_c$ and $\lambda_r$) that are applied to the filter input. Key events that occur during the automatic tuning process are labeled in  \textbf{Fig. \ref{fig:dac_sampling}} and explained in \textbf{Table \ref{tab:dac_timestamps}}. This figure shows three iterations to tune the filter (rings, Q-PS \& B-C), and the time taken for the iteration progressively decreases as the filter gets closer to the thermal steady-state during tuning. In our filter layout, the Q-PS and B-C coupler were in close proximity to the rings resulting in substantial thermal crosstalk. The fabricated $2^{nd}$-order filter takes around 725 seconds to tune for the first time. The long configuration times are primarily determined by the thermal crosstalk between the components and can be significantly reduced by spreading out the layout and by employing microheaters with undercut for thermal isolation \cite{masood2013comparison,coenen2022thermal}.

\begin{table}[htbp]
\centering
\caption{Explanation of DAC tuning events.}
\begin{tabular}{p{0.06\linewidth}p{0.86\linewidth}}
\hline
Event & Details\\
\hline
$A$ & Upper Ring coupler (UR-C) is set to zero coupling after a coarse and fine search.\\
\hline
$B$ & Lower Ring coupler (LR-C) is set to zero coupling after a coarse and fine search.\\
\hline
$C$ & Upper Ring coupler (UR-C) is set to desired coupling ratio based on \textbf{Algorithm \ref{alg:ccoeffp2}}.\\
\hline
$D$ & Lower Ring coupler (LR-C) is set to desired coupling ratio based on \textbf{Algorithm \ref{alg:ccoeffp2}}.\\
\hline
$E_n$ & $n^{th}$ iteration of Upper Ring phase-shifter (UR-PS) resonance alignment to $\lambda_c$ after coarse and fine search. \\
\hline
$F_n$ & $n^{th}$ iteration of Lower Ring phase-shifter (LR-PS) resonance alignment to $\lambda_c$ after coarse and fine search.\\
\hline
$G_n$ & $n^{th}$ iteration of configuring Upper Ring phase-shifter (UR-PS) phase bias $\phi_{1}$ based on \textbf{Algorithm \ref{alg:ringbiasp2}}.\\
\hline
$H_n$ & $n^{th}$ iteration of configuring Lower Ring phase-shifter (LR-PS) phase bias $\phi_{2}$ based on \textbf{Algorithm \ref{alg:ringbiasp2}}.\\
\hline
$I_n$ & $n^{th}$ iteration of maximizing 10$\%$ monitor tap output by tuning Quadrature phase-shifter (Q-PS).\\
\hline
$J_n$ & $n^{th}$ iteration of maximizing 10$\%$ monitor tap output by tuning Back-coupler (B-C).\\
\hline
\end{tabular}
  \label{tab:dac_timestamps}
\end{table}

\begin{figure}[htbp]
\centering
\includegraphics[width=1\linewidth]{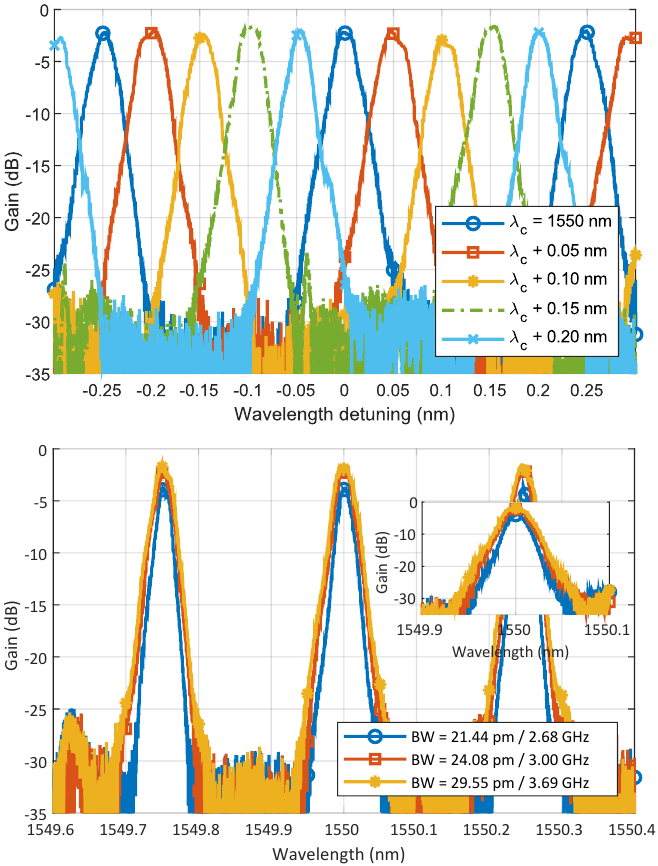}
\caption{(Top) Center frequency tuning of the silicon photonic filter. Here, the filter is configured at 5 different wavelengths with $\Delta\lambda = 0.05$ nm or $\Delta f = 6.25$ GHz. (bottom) Bandwidth Tuning of the filter around $\lambda_c = 1550$ nm.}
\label{fig:fc_bw_tuning}
\end{figure}

As mentioned earlier, for software defined radios (SDR) and flexible RF photonic receivers, it is vital to have filters with agile center-frequency and bandwidth tunability. To demonstrate such capability, we reconfigured the filter at 5 different center frequencies ($\lambda_c$) with 0.05nm ($\sim$6.25 GHz) spacing as shown in \textbf{Fig. \ref{fig:fc_bw_tuning}} (top). It's important to note that the center frequency tuning is continuous and any center wavelength ($\lambda_c$) can be configured as long as it is within the FSR of the rings. This means that the fabricated filter can be reconfigured at any frequency between DC and 31 GHz. To cover higher frequency bands, the filter can be redesigned with smaller rings (i.e. larger FSR). On the other hand, the bandwidth (BW) tunability of our proposed filter is shown in \textbf{Fig. \ref{fig:fc_bw_tuning}} (bottom) for BW = 2.68, 3 and 3.69 GHz. A trade-off can be observed in \textbf{Fig. \ref{fig:fc_bw_tuning}} (bottom), whereby as the BW is reduced, the passband insertion loss increases \cite{rasras2007demonstration}. This becomes even more prominent when sub-GHz 3dB bandwidth settings are attempted. This is due to the losses in the rings, and can be mitigated by adopting low-loss multimode waveguide design \cite{onural2020ultra} and/or utilizing the upcoming ultra low-loss PDK from AIM Photonics \cite{aimphot}.

In \textbf{Table \ref{tab:performance_comparison}}, a comprehensive comparison of this work with other related work such as $2^{nd}$-order APD filter \cite{choo2018automatic}, $4^{th}$-order APD filter \cite{choo2018automatic}, $5^{th}$-order CROW filter \cite{mak2015automatic}, $2^{nd}$-order CROW filter \cite{mak2016programmable}, $4^{th}$-order APD filter \cite{rasras2007demonstration}, Benes switch matrix based filter \cite{shen2019silicon}, filter based on phase-to-intensity modulation conversion \cite{zhang2018chip} and higher order vernier filter \cite{jayatilleka2017automatic} is provided. As evident from the comparison, this work presents a first fully-automatic filter tuning scheme with in-situ reconfiguration capability and a significant step towards a portable solution. 

\begin{table*}[t]
\scriptsize
\centering
\caption{Comparison with state-of-the art in filter tuning.}
\begin{tabular}{|cc|c|c|c|c|c|c|c|c|c|}
\hline
\multicolumn{2}{|c|}{\textbf{Metric}}                                                                                           & \begin{tabular}[c]{@{}c@{}}This\\ Work\end{tabular}      & \begin{tabular}[c]{@{}c@{}}Ref.\\ \cite{choo2018automatic}\end{tabular}    & \begin{tabular}[c]{@{}c@{}}Ref.\\ \cite{choo2018automatic}\end{tabular}    & \begin{tabular}[c]{@{}c@{}}Ref.\\ \cite{mak2015automatic}\end{tabular}          & \begin{tabular}[c]{@{}c@{}}Ref.\\ \cite{mak2016programmable}\end{tabular}    & \begin{tabular}[c]{@{}c@{}}Ref.\\ \cite{rasras2007demonstration}\end{tabular}       & \begin{tabular}[c]{@{}c@{}}Ref.\\ \cite{shen2019silicon}\end{tabular}    & \begin{tabular}[c]{@{}c@{}}Ref.\\ \cite{zhang2018chip}\end{tabular} & \begin{tabular}[c]{@{}c@{}}Ref.\\ \cite{jayatilleka2017automatic}\end{tabular}      \\ \hline \hline
\multicolumn{2}{|c|}{\textbf{Technology}}                                                                                       & AIM                                                      & \begin{tabular}[c]{@{}c@{}}IME\\ SiP\end{tabular}        & \begin{tabular}[c]{@{}c@{}}IME\\ SiP\end{tabular}        & \begin{tabular}[c]{@{}c@{}}IME\\ SiP\end{tabular}              & \begin{tabular}[c]{@{}c@{}}IME\\ SiP\end{tabular}        & \begin{tabular}[c]{@{}c@{}}BAE\\ SiP\end{tabular}           & SiP                                                      & SiP                                                   & \begin{tabular}[c]{@{}c@{}}IME\\ SiP\end{tabular}          \\ \hline
\multicolumn{2}{|c|}{\textbf{Architecture}}                                                                                     & APD                                                      & APD                                                      & APD                                                      & CROW                                                           & CROW                                                     & APD                                                         & Benes                                                    & PM-IM                                                 & Vernier                                                    \\ \hline
\multicolumn{2}{|c|}{\textbf{Order}}                                                                                            & 2                                                        & 2                                                        & 4                                                        & 5                                                              & 2                                                        & 4                                                           & 2, 4                                                     & -                                                     & 4                                                          \\ \hline
\multicolumn{2}{|c|}{\textbf{FSR (nm)}}                                                                                         & \begin{tabular}[c]{@{}c@{}}0.249\\ (31 GHz)\end{tabular} & \begin{tabular}[c]{@{}c@{}}0.396\\ (50 GHz)\end{tabular} & \begin{tabular}[c]{@{}c@{}}0.396\\ (50 GHz)\end{tabular} & \begin{tabular}[c]{@{}c@{}}$\sim$2.47\\ (308 GHz)\end{tabular} & \begin{tabular}[c]{@{}c@{}}2.10\\ (265 GHz)\end{tabular} & \begin{tabular}[c]{@{}c@{}}0.1307\\ (16.5 GHz)\end{tabular} & \begin{tabular}[c]{@{}c@{}}8.3\\ (1.03 THz)\end{tabular} & -                                                     & \begin{tabular}[c]{@{}c@{}}37.21\\ (4.64 THz)\end{tabular} \\ \hline
\multicolumn{2}{|c|}{\textbf{BW (GHz)}}                                                                                         & 3                                                        & 6.89                                                     & 5                                                        & $\sim$30.9                                                     & 33.4                                                     & 1                                                           & $\sim$46, $\sim$56                                       & 1.93                                                  & 39                                                         \\ \hline
\multicolumn{2}{|c|}{\textbf{Rejection (dB)}}                                                                                   & 30                                                       & 32                                                       & 33                                                       & $\sim$50                                                       & 25                                                       & 25                                                          & 16, 23.4                                                 & 18                                                    & 50                                                         \\ \hline
\multicolumn{2}{|c|}{\textbf{Passband Insertion Loss (dB)}}                                                                     & 2.2                                                      & $\sim$2.3                                                & 4.67                                                     & $\sim$2.64                                                     & 2.11                                                     & 6                                                           & 10.3, 11.2                                               & 38.9                                                  & 4.5                                                       \\ \hline
\multicolumn{1}{|c|}{\multirow{3}{*}{\textbf{Reconfigurability}}}                                                   & $f_c$        & Y                                                        & Y                                                        & Y                                                        & Y                                                              & Y                                                        & Y                                                           & Y                                                        & Y                                                     & Y                                                          \\ \cline{2-11} 
\multicolumn{1}{|c|}{}                                                                                     & BW        & Y                                                        & Y                                                        & Y                                                        & N                                                              & Y                                                        & Y                                                           & Y                                                        & N                                                     & N                                                          \\ \cline{2-11} 
\multicolumn{1}{|c|}{}                                                                                     & Rejection & Y                                                        & Y                                                        & Y                                                        & N                                                              & N                                                        & Y                                                           & N                                                        & N                                                     & N                                                          \\ \hline
\multicolumn{1}{|c|}{\multirow{3}{*}{\begin{tabular}[c]{@{}c@{}}\textbf{Automatic}\\ \textbf{Reconfiguration}\end{tabular}}} & $f_c$        & Y                                                        & Y                                                        & Y                                                        & Y                                                              & Y                                                        & N                                                           & N                                                        & N                                                     & Y                                                          \\ \cline{2-11} 
\multicolumn{1}{|c|}{}                                                                                     & BW        & Y                                                        & Y                                                        & Y                                                        & N                                                              & Y                                                        & N                                                           & N                                                        & N                                                     & N                                                          \\ \cline{2-11} 
\multicolumn{1}{|c|}{}                                                                                     & Rejection & Y                                                        & Y                                                        & Y                                                        & N                                                              & N                                                        & N                                                           & N                                                        & N                                                     & N                                                          \\ \hline
\multicolumn{2}{|c|}{\textbf{Portability}}                                                                                      & \begin{tabular}[c]{@{}c@{}}No\\ OVNA\end{tabular}        & \begin{tabular}[c]{@{}c@{}}Uses\\ OVNA\end{tabular}      & \begin{tabular}[c]{@{}c@{}}Uses\\ OVNA\end{tabular}      & -                                                              & -                                                        & -                                                           & -                                                        & -                                                     & -                                                          \\ \hline
\end{tabular}
  \label{tab:performance_comparison}
\end{table*}

\section{Conclusion and Future Work}

This work presented a robust silicon photonic filter for RF photonic applications with an improved reconfiguration algorithm that is insensitive to PVT variations. By utilizing proposed in-situ configuration algorithm with simpler pre-characterization steps (i.e. without requiring an OVNA), we have experimentally shown that the filter can be configured to any desired center frequency, bandwidth and rejection as per the user specifications with high fidelity. Moreover, the filter is a first of its class to be fabricated in AIM photonics's CMOS compatible Active Photonic process, making the filter widely manufacturable at low cost for next-generation RF photonic systems. The proposed coupling coefficient extraction and configuration algorithm provides upto 2X improvement in configuration time over the prior art.

Further improvements such as sub-GHz bandwidth with low insertion loss can potentially be achieved by incorporating multimode $\&$ low-loss waveguide designs \cite{onural2020ultra} and $<$1$\%$ monitor taps in the rings, and/or leveraging the planned ultra-low-loss PDK and MPW runs from AIM Photonics \cite{aimphot}. The current $2^{nd}$-order filter initial reconfiguration time is around 725s. With our optimized algorithm and a low thermal crosstalk design where back-coupler and quadrature phase shifter tuning does not affect the ring bias ($\phi_n$) as in \cite{choo2018automatic}, the filter reconfiguration time can be as fast as $\sim$300s. This can be achieved by careful planning of PIC layout where the  thermo-optic phase-shifters (microheaters) are placed further apart from sensitive waveguides (rings) at the expense of a larger  footprint. Achieving rapid filter reconfiguration on the order of seconds in frequency-agile SDRs will require even higher thermal isolation between on-chip photonic components. This can potentially be achieved by integration of microheaters with undercut \cite{coenen2022thermal} in the SiP foundry process.  In summary, by taking full advantage of the large-volume low-cost manufacturing of a CMOS-compatible photonic process, the presented RF photonic filters and improved reconfiguration algorithm provide a way forward for wider adoption in high-performance RF and microwave photonic transceivers.

\section{Acknowledgment}
The authors gratefully acknowledge the generous funding support from the Air Force Office of Sponsored Research (AFOSR) YIP Award FA9550-17-1-0076 and DARPA YFA Award HR00112110001.

\ifCLASSOPTIONcaptionsoff
  \newpage
\fi


\bibliographystyle{IEEEtran}
\bibliography{filter}

\end{document}